\documentclass[%
 reprint,
superscriptaddress,
 amsmath,amssymb,
 aps,
 article
]{revtex4-1}

\usepackage{graphicx}
\usepackage{dcolumn}
\usepackage{bm}
\usepackage{textcomp}
\usepackage{soul}
\usepackage[normalem]{ulem}
\usepackage{subfigure}
\usepackage{array}
\usepackage{diagbox}
\usepackage{xcolor}
\relax

\begin{document}

\preprint{APS/123-QED}

\title{Network Models for Characterization of Trabecular Bone}
\author{Avik Mondal}
\altaffiliation[Current affiliation: ]{Department of Physics, University of Michigan, Ann Arbor, MI 48109}
\affiliation{Department of Physics, University of California, Santa Barbara, Santa Barbara, CA 93106, USA
}%
\author{Chantal Nguyen} 
\email{cnguyen@physics.ucsb.edu}
\affiliation{Department of Physics, University of California, Santa Barbara, Santa Barbara, CA 93106, USA
}%
\author{Xiao Ma}
\author{Ahmed E. Elbanna}
\affiliation{Department of Civil and Environmental Engineering, University of Illinois at Urbana-Champaign, Urbana, IL 61801, USA
}%
\author{Jean M. Carlson}
\affiliation{Department of Physics, University of California, Santa Barbara, Santa Barbara, CA 93106, USA
}%

\date{\today}

\begin{abstract}
Trabecular bone is a lightweight, compliant material organized as a web of struts and rods (trabeculae) that erode with age and the onset of bone diseases like osteoporosis, leading to increased fracture risk.  The traditional diagnostic marker of osteoporosis, bone mineral density (BMD), has been shown in \textit{ex vivo} experiments to correlate poorly with fracture resistance when considered on its own, while structural features in conjunction with BMD can explain more of the variation in trabecular bone strength. We develop a network-based model of trabecular bone by creating graphs from micro-CT images of human bone, with weighted links representing trabeculae and nodes representing branch points. These graphs enable calculation of quantitative network metrics to characterize trabecular structure. We also create finite element models of the networks in which each link is represented by a beam, facilitating analysis of the mechanical response of the bone samples to simulated loading. We examine the structural and mechanical properties of trabecular bone at the scale of individual trabeculae (of order 0.1 mm) and at the scale of selected volumes of interest (approximately a few mm), referred to as VOIs. At the VOI scale, we find significant correlations between the stiffness of VOIs and ten different structural metrics. Individually, the volume fraction of each VOI is most strongly correlated to the stiffness of the VOI. We use multiple linear regression to identify the smallest subset of variables needed to capture the variation in stiffness. In a linear fit, we find that {node degree, weighted node degree, Z-orientation, weighted Z-orientation, trabecular spacing, link length, and the number of links} are the structural metrics that are most significant ($p$ $< 0.05$) in capturing the variation of stiffness in trabecular networks.

\end{abstract}

\pacs{Valid PACS appear here}
\maketitle
\frenchspacing
\section{\label{sec:intro} Introduction}
Trabecular bone is a porous, web-like arrangement of bone struts and rods (trabeculae), resulting in a strong yet lightweight and flexible tissue. One of two types of bone in the body, trabecular bone is found primarily in the vertebrae, wrist, hip, and femur, encased within a stiff shell of cortical bone, which is the other type of bone. Trabeculae erode with age; this process is accelerated with bone disease. Osteoporosis is a systemic skeletal disease characterized by low bone mass and micro-architectural deterioration of bone tissue, leading to fragility and increased susceptibility to fracture. It is estimated that osteoporosis affects approximately 10.2 million adults in the United States aged fifty years or older \cite{mcdonnel_et_al}, in addition to millions worldwide \cite{osteoporosis_conference}. Each year, an estimated 1.5 million Americans experience a fracture due to bone disease \cite{surgeon_general}. Hip fracture is associated with a 20\% excess mortality in the year following the fracture \cite{low_bmd_postmenopausal_women}. In 1995, the cost of managing fractures was approximately \$13.8 billion dollars in the United States alone and is projected to increase as life expectancy increases \cite{low_bmd_postmenopausal_women}.

Currently, estimation of areal bone mineral density (BMD) is the conventional method for diagnosis of osteoporosis and prediction of fracture risk \cite{boutroy_invivo_qct}. The two most widely used methods of estimating BMD are dual X-ray absorptiometry (DXA), which measures density via the attenuation of x-rays by bone at different energies \cite{dual_xray_absorptiometry}, and quantitative computed tomography (QCT), in which bone density is calculated from low-resolution 2-D image slices of bone \cite{surgeon_general}. However, recent studies suggest that BMD alone is a poor indicator of bone strength. On its own, it has been reported to account for between 40\% and 70\% of the variation in the compressive yield strength of trabecular bone \cite{mcdonnel_et_al,elastic_properties_fabric_tensor, ding_et_al,silva_paper}, while taking both BMD and trabecular architecture into account can reportedly explain up to 90\% of the variance in bone strength as measured in \textit{ex vivo} mechanical tests \cite{mcdonnel_et_al,elastic_properties_fabric_tensor, goulet_paper, trab_micro_structure_biomech_behavior}. 

The architecture of trabecular bone is typically characterized with bone histomorphometry, the image-analysis-based study of bone tissue to obtain quantitative information about bone structure and remodeling \cite{histo_source,histo_source2}. Modern histomorphometry is accomplished using high-resolution imaging, such as micro-CT ({\textmu}CT), which can capture image resolution down to the order of a micron \cite{guidelines_assessment, HighResMicroCT}. However, the large amounts of radiation involved in high-resolution tomography limits its \textit{in vivo} usage to distal extremities in humans \cite{hrpqct}. In this study, we utilize high-resolution {\textmu}CT images of cadaveric vertebral bone, from which we generate accurate 3-D reconstructions of trabecular volumes and extract histomorphometric parameters.

The web-like structure of trabecular bone closely resembles a network, i.e., a system of nodes, or vertices, that are connected by links, or edges. Each trabecula resembles a link, while the points at which multiple trabeculae meet, referred to here as branch points, resemble nodes. Hence, we capitalize on this resemblance by modeling trabecular bone as a network. We exploit the existing mathematical framework developed in network science to analyze the topology of trabecular bone in a streamlined fashion. Network science has rarely been applied to the study of bone \cite{tommasini_mice} but has been used to study a variety of systems across disciplines, including social and ecological systems, biological vasculature, granular materials, and soil \cite{newman_book,hu_cai,particles_grains,soil_networks}.

We begin by converting {\textmu}CT images of trabecular bone into network models that are compactly represented in a mathematical form, in contrast to previous methods of trabecular analysis that involve specialized image processing techniques \cite{boutroy_invivo_qct, trab_micro_structure_biomech_behavior}.
To relate structure to mechanics, we also create two types of finite element models that respectively correspond to 3-D realizations of the bone images and of the network models.

We examine the statistical variability in architectural and mechanical properties across scales. At the smallest scales, we characterize individual trabeculae and branch points with network metrics. Moreover, we compute distributions of these metrics for a network derived from a volume of bone that may contain hundreds of trabeculae and branch points. For a mesoscale analysis, we coarse-grain an entire vertebral body into such volumes (Fig. \ref{fig:VOI_coarse_grain}) and compare distributions across these volumes. 

Likewise, we analyze mechanical response across scales with simulated deformation of the bone models. The stress in one trabecula -- here modeled as a beam in a finite element model -- represents the smallest-scale mechanical measure, and mesoscale response is represented by the overall stiffness of bone volumes. We compare the stress distribution of a beam network and its stiffness with structural metrics. Analysis at the mesoscale reveals several correlations between architectural and mechanical quantities in bone. 

\section{\label{sec: background} Methods}
\subsection{{\textmu}CT image analysis}

To develop network models of trabecular bone, we utilize a 37 {\textmu}m resolution {\textmu}CT image set obtained from the Bone 3D Project Team \cite{bone_data}.
This set includes 970 axial image slices, each 2048 pixel $\times$ 2048 pixel (75 mm $\times$ 75 mm) in size, of vertebral body L3 from a human cadaver, imaged using the Scanco {\textmu}CT 80 scanner. Stacked along the axial direction, the images encompass a volume with dimensions 75 mm $\times$ 75 mm $\times$ $35.9$ mm.

\begin{figure*}[t]
\centering
\includegraphics[width=\linewidth]{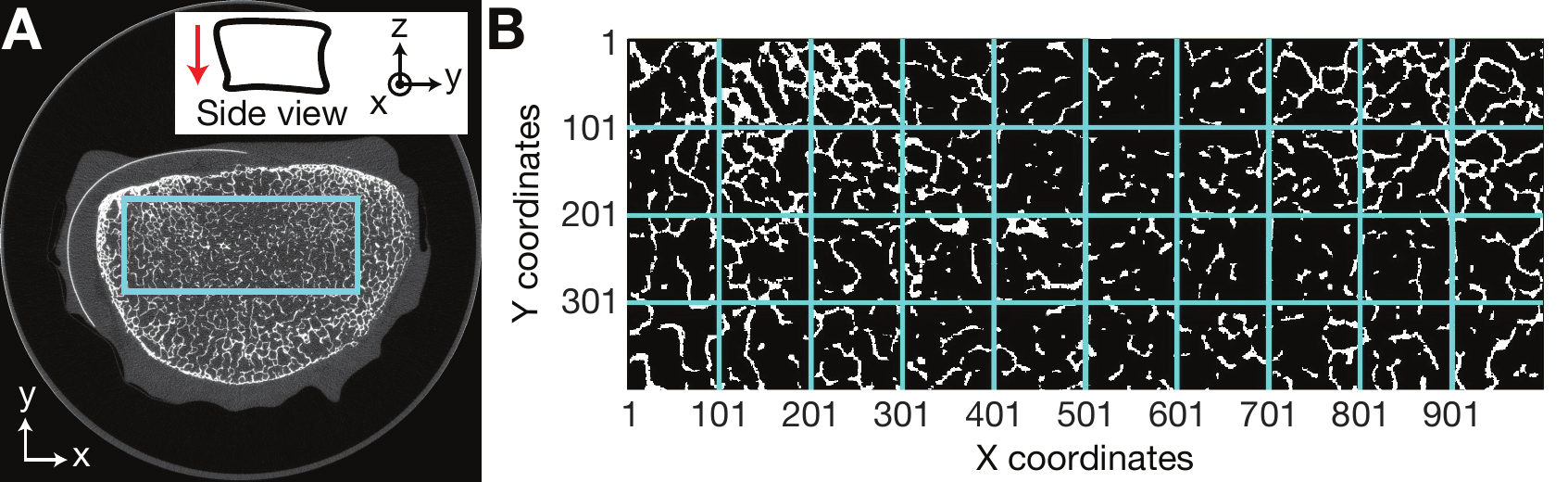}
\caption{\label{fig:VOI_coarse_grain} Trabecular bone images used in this study. A: {\textmu}CT transverse image slice of human vertebral body L3 \cite{bone_data}. The highlighted region is divided into volumes of interest (VOIs) shown in B. {The inset shows a schematic of a sagittal cross-section of a human vertebral body as it corresponds to our sample. The red arrow indicates the principal direction of loading.} B: The selected region is divided into 100 pixel (3.7 mm) $\times$ 100 pixel tiles; each tile shown is the top image of a 100-image stack (Z-coordinate) that defines a VOI. The X, Y, and Z directions refer to the medial-lateral, anterior-posterior, and superior-inferior directions, respectively.}
\end{figure*}

Pre-processing of {\textmu}CT images is performed with CT-analyser (CTAn) \cite{CTAn}.
The raw images are binarized using the Otsu thresholding method \cite{otsu}. All the images undergo a ``despeckling'' procedure to remove spurious pixels; all black or white clusters consisting of fewer than 100 pixels in three dimensions are removed. The stack is divided into small volumes of interest (VOIs) to facilitate future processing. Each VOI comprises a stack of 100 images that are each 100 pixel $\times$ 100 pixel, corresponding to a cube with dimensions (3.7 mm)$^3$, or a volume of approximately 50 mm$^3$ (Fig. \ref{fig:VOI_coarse_grain}).

\subsection{Generating networks} 
\begin{figure}[h]
\centering
\includegraphics[width=\linewidth]{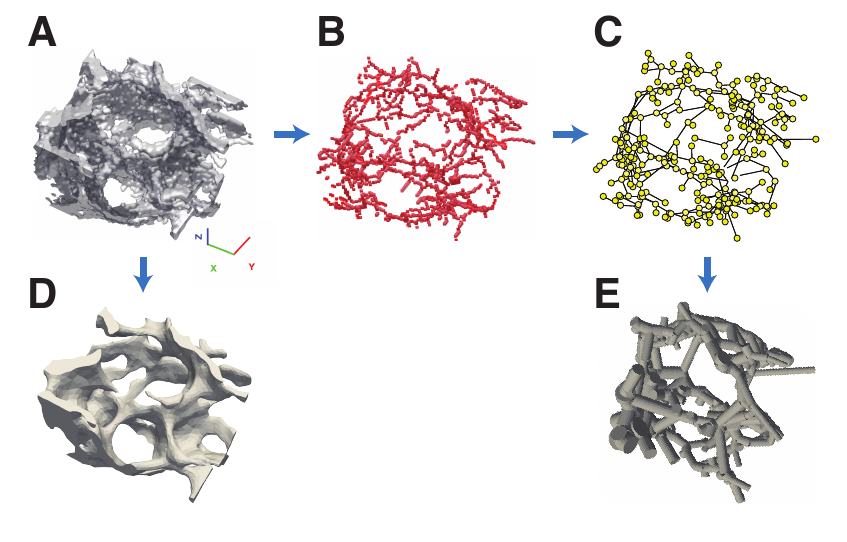}
\caption{\label{fig:samplepipeline} Trabecular bone modeling pipeline. {\textmu}CT image slices of bone are stacked to create a 3-D volume (A). The skeleton (B) is generated by iteratively thinning the volume until a one-voxel-wide line remains. Branch points in the skeleton are assigned as nodes (yellow circles) in a network (C), with edges representing trabeculae connecting the branch points. Endpoints of trabeculae created as a result of image segmentation are also assigned as nodes. A continuum finite element model (D) is generated by meshing the original bone images, while a beam-element model (E) is generated by converting each edge in the network to a beam, where the thickness of the beam is defined relative to the edge weight.}
\end{figure}

Network models of trabecular bone are derived using skeletonization, a process that isolates the medial axis of an image -- the ``skeleton'' \cite{lee_skeletonization}. The medial axis of an object in 3-D is the locus of the centers of the maximally fitting spheres where the spheres touch the surface of the object at more than one point. 
Due to the web-like structure of trabecular bone and the rod-like geometry of individual trabeculae, the medial axis of a section of trabecular bone is usually a collection of connected lines, each running through what previously was the center of each of the trabeculae (Fig. \ref{fig:samplepipeline}). 

We use the Skeleton3D library \cite{osteocyte_kerschnitzki} for MATLAB (MathWorks, Natick, MA) to compute trabecular skeletons for each VOI. This library utilizes an algorithm that skeletonizes an image by iteratively removing surface voxels from the volume in such a way that the topology of the sample is preserved. As a result, all branch points and cavities in the original shape remain after each iteration. This process is repeated until all that remains is a collection of one-voxel-thick segments \cite{osteocyte_kerschnitzki}.

The Skel2Graph library \cite{osteocyte_kerschnitzki} for MATLAB is used to convert the skeletons into networks. Links are defined as individual trabeculae, and nodes as the branch points between trabeculae. The process of dividing the bone into VOIs results in isolated trabeculae in each VOI that ``float'' in space; these are removed, and a single connected component is isolated.
The links are weighted with the average thicknesses of the individual trabeculae. Bone thickness is computed with the BoneJ plugin \cite{bonej} for Fiji \cite{FIJI}, a biological image-focused distribution of ImageJ (National Institutes of Health, Bethesda, MD).

\section{\label{sec: results} Results}
\subsection{\label{topologicalnetworkanalysis} Structural analysis}

\begin{figure*}[tbp]
\centering
\includegraphics[width=\textwidth]{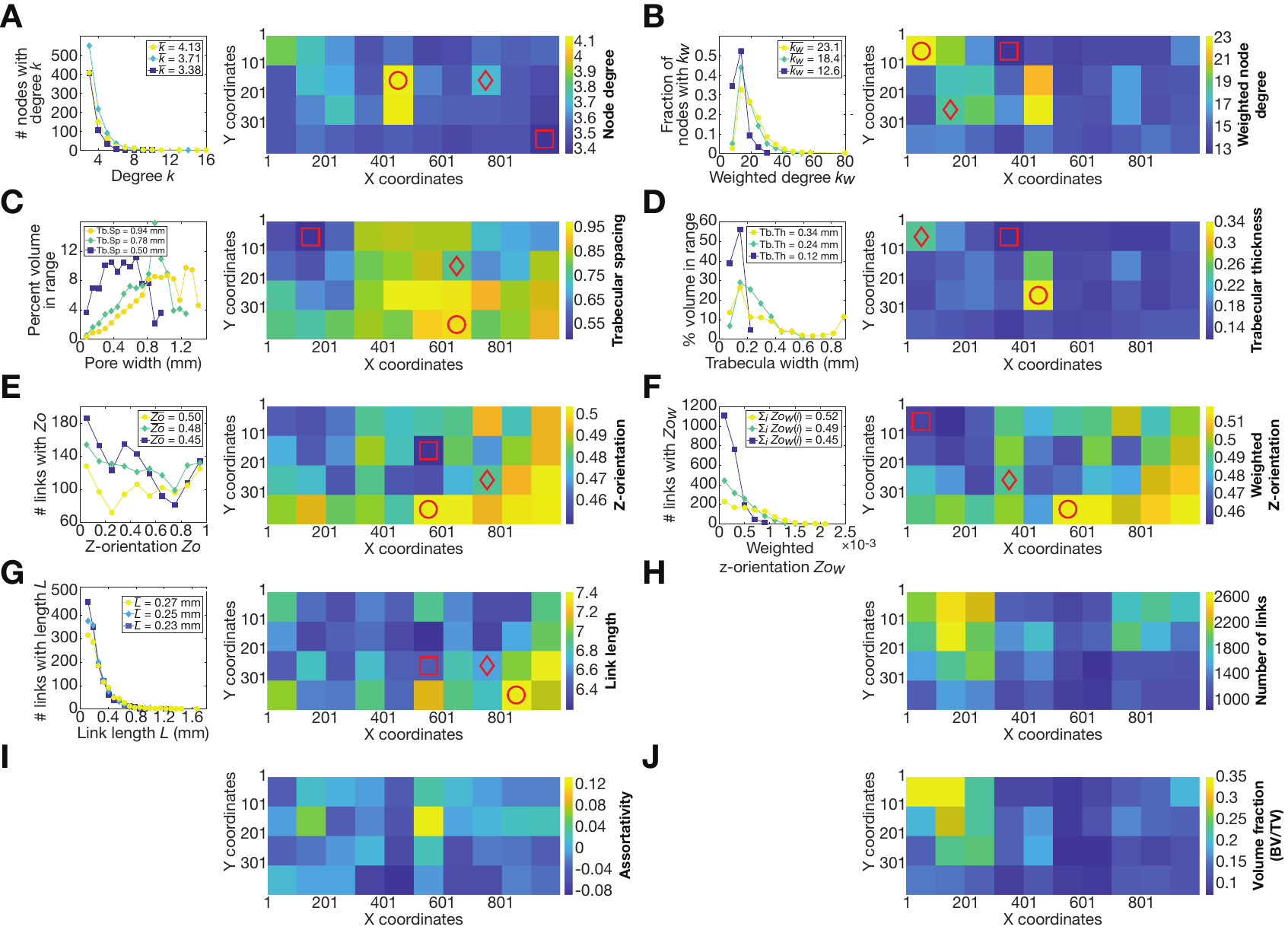}
\caption{\label{fig:multiscale_analysis} 
Distributions of structural metrics. A: node degree; B: weighted node degree; C: trabecular spacing (Tb.Sp); D: trabecular thickness (Tb.Th); E: Z-orientation; F: weighted Z-orientation; G: link length; H: number of links; I: assortativity; J: volume fraction (BV/TV). Each panel consists of two plots, except for panels H, I, and J: the left plot illustrates the distribution of metrics at the node/link scale, and the right plot shows the distribution of metrics at the VOI scale. (Number of links (H), assortativity (I) and volume fraction (J) are only defined at the VOI scale.) The node/link-scale plots show distributions within three example VOIs; the mean (or sum, in the case of weighted Z-orientation) of each distribution is indicated in the respective top right corners. Values are binned, with markers indicating the midpoint of each bin, except for node degree, which takes integer values. The VOI-scale plots illustrate the spatial distributions of structural metrics across the vertebral body. The color of each tile represents the average structural metric for one VOI. The three VOIs for which the histograms are plotted on the left are indicated on the right by shapes corresponding to their respective markers and illustrate results for representative high (yellow circles), mid-range (light green/blue diamonds), and low (dark blue squares) values of the corresponding VOI scale metrics.}
\end{figure*}

We characterize the structure of bone by investigating histomorphometry, geometry, and network topology at the scale of individual trabeculae (of order 0.1 mm) and at the VOI scale (approximately a few mm). At the smaller scale, we determine characteristics of nodes and links, as well as their distributions within a VOI. At the larger scale, we compare the distributions of such characteristics across VOIs and examine the spatial distribution of structural properties (Fig. \ref{fig:multiscale_analysis}). We analyze a total of 40 VOIs, each measuring (3.7 mm)$^3 \approx 50$ mm$^3$. Each VOI is small enough for structural and mechanical properties to be calculated in a short amount of computational time while large enough to capture significant structural variation.

\begin{table*}
\def\arraystretch{1.5}
\setlength\tabcolsep{3pt}
\centering
\begin{tabular}{|c||c|c|c|c|c|c|c|c|c|c|} 
\hline
Metric                                                                                        & \multicolumn{1}{l|}{$A$ } & $k$                                                                         & $k_w$                                                                       & BV/TV                                                                       & Tb.Sp                                                                        & Tb.Th                                                                       & $L$                                                                         & $Zo$                                                                           & $Zo_w$                                                              & No. links                                                                                   \\ 
\hline\hline
Assortativity ($A$)                                                                           & --                         & \begin{tabular}[c]{@{}c@{}}\textbf{0.7219},\\ \textbf{ 0.001} \end{tabular} & \begin{tabular}[c]{@{}c@{}}\textbf{0.6959},\\ \textbf{ 0.001} \end{tabular} & \begin{tabular}[c]{@{}c@{}}\textbf{0.5610},\\ \textbf{ 0.001} \end{tabular} & \begin{tabular}[c]{@{}c@{}}\textbf{-0.3207},\\ \textbf{0.0437} \end{tabular} & \begin{tabular}[c]{@{}c@{}}\textbf{0.7260},\\ \textbf{ 0.001} \end{tabular} & \begin{tabular}[c]{@{}c@{}}\textbf{0.6279},\\ \textbf{ 0.001} \end{tabular} & \begin{tabular}[c]{@{}c@{}}\textbf{-0.3647},\\ \textbf{0.0207} \end{tabular} & \begin{tabular}[c]{@{}c@{}}\textbf{-0.4550},\\\textbf{0.0032}\end{tabular} & {\begin{tabular}[c]{@{}c@{}}\textbf{0.4757},\\\textbf{0.0019}\end{tabular}}  \\ 
\hline
Node degree ($k$)                                                                             & --                         & --                                                                           & \begin{tabular}[c]{@{}c@{}}{\textbf{0.8651},}\\ {\textbf{ \textless{}0.001}} \end{tabular} & \begin{tabular}[c]{@{}c@{}}{\textbf{0.5715},}\\ {\textbf{\textless{} 0.001}} \end{tabular} & \begin{tabular}[c]{@{}c@{}}{-0.2712,}\\ { 0.091 }\end{tabular} & \begin{tabular}[c]{@{}c@{}}{\textbf{0.8730},}\\ {\textbf{\textless{} 0.001}} \end{tabular} & \begin{tabular}[c]{@{}c@{}}{\textbf{0.6925},}\\ {\textbf{\textless{} 0.001}} \end{tabular} & \begin{tabular}[c]{@{}c@{}}{\textbf{-0.6229},}\\ {\textbf{\textless{} 0.001}} \end{tabular}                   & \begin{tabular}[c]{@{}c@{}}{\textbf{-0.7077},}\\{\textbf{\textless{} 0.001}}\end{tabular} & \begin{tabular}[c]{@{}c@{}}{\textbf{0.5208},}\\{\textbf{ \textless{} 0.001}}\end{tabular}                        \\ 
\hline
\begin{tabular}[c]{@{}c@{}}Weighted node\\ degree ($k_w$) \end{tabular}                       & --                         & --                                                                           & --                                                                           & \begin{tabular}[c]{@{}c@{}}\textbf{0.7575},\\\textbf{ \textless{}0.001} \end{tabular}  & \begin{tabular}[c]{@{}c@{}}\textbf{-0.4083},\\ \textbf{0.0089} \end{tabular} & \begin{tabular}[c]{@{}c@{}}\textbf{0.9051},\\ \textbf{\textless{} 0.001} \end{tabular} & \begin{tabular}[c]{@{}c@{}}\textbf{0.7026},\\ \textbf{\textless{} 0.001} \end{tabular} & \begin{tabular}[c]{@{}c@{}}\textbf{-0.5411},\\ \textbf{\textless{} 0.001} \end{tabular} & \begin{tabular}[c]{@{}c@{}}\textbf{-0.6593},\\\textbf{\textless{}0.001}\end{tabular}   & \begin{tabular}[c]{@{}c@{}}\textbf{0.5686},\\\textbf{ \textless{} 0.001}\end{tabular}                        \\ 
\hline
\begin{tabular}[c]{@{}c@{}}Volume fraction\\ (BV/TV) \end{tabular}                            & --                         & --                                                                           & --                                                                           & --                                                                           & \begin{tabular}[c]{@{}c@{}}\textbf{-0.8193},\\ \textbf{\textless{} 0.001} \end{tabular} & \begin{tabular}[c]{@{}c@{}}\textbf{0.5040},\\\textbf{ \textless{}0.001} \end{tabular}  & \begin{tabular}[c]{@{}c@{}}0.2792,\\ 0.0810 \end{tabular}                   & \begin{tabular}[c]{@{}c@{}}\textbf{-0.4630},\\ \textbf{0.0026} \end{tabular} & \begin{tabular}[c]{@{}c@{}}\textbf{-0.6022},\\\textbf{\textless{}0.001}\end{tabular}  & \begin{tabular}[c]{@{}c@{}}\textbf{0.8673},\\\textbf{\textless{}0.001}\end{tabular}                        \\ 
\hline
Tb.Sp                                                                                         & --                         & --                                                                           & --                                                                           & --                                                                           & --                                                                            & \begin{tabular}[c]{@{}c@{}}-0.1008,\\ 0.5358 \end{tabular}                  & \begin{tabular}[c]{@{}c@{}}0.1522,\\ 0.3484 \end{tabular}                   & \begin{tabular}[c]{@{}c@{}}\textbf{0.3169},\\ \textbf{0.0463} \end{tabular}  & \begin{tabular}[c]{@{}c@{}}\textbf{0.4665},\\\textbf{0.0024}\end{tabular}   & \begin{tabular}[c]{@{}c@{}}\textbf{-0.9047},\\\textbf{\textless{}0.001}\end{tabular}                       \\ 
\hline
Tb.Th                                                                                         & --                         & --                                                                           & --                                                                           & --                                                                           & --                                                                            & --                                                                           & \begin{tabular}[c]{@{}c@{}}\textbf{0.8672},\\ \textbf{\textless{}0.001} \end{tabular} & \begin{tabular}[c]{@{}c@{}}\textbf{-0.4492},\\\textbf{ 0.0036} \end{tabular} & \begin{tabular}[c]{@{}c@{}}\textbf{-0.5229},\\\textbf{\textless{}0.001}\end{tabular}   & \begin{tabular}[c]{@{}c@{}}\textbf{0.3225},\\\textbf{0.0424}\end{tabular}                       \\ 
\hline
\begin{tabular}[c]{@{}c@{}}Link length\\ ($L$) \end{tabular}                                  & --                         & --                                                                           & --                                                                           & --                                                                           & --                                                                            & --                                                                           & --                                                                           & \begin{tabular}[c]{@{}c@{}}-0.1569,\\0.3337 \end{tabular}                    & \begin{tabular}[c]{@{}c@{}}-0.2063,\\0.2016\end{tabular}           & \begin{tabular}[c]{@{}c@{}}0.0168,\\0.9180\end{tabular}                                \\ 
\hline
\begin{tabular}[c]{@{}c@{}}Z-orientation\\ ($Zo$) \end{tabular}                                 & --                         & --                                                                           & --                                                                           & --                                                                           & --                                                                            & --                                                                           & --                                                                           & --                                                                            & \begin{tabular}[c]{@{}c@{}}\textbf{0.9626},\\\textbf{\textless{}0.001}\end{tabular}    & \begin{tabular}[c]{@{}c@{}}\textbf{-0.4578},\\\textbf{0.0030}\end{tabular}                      \\ 
\hline
{\begin{tabular}[c]{@{}c@{}}Weighted Z-orientation\\($Zo_w$)\end{tabular}} & --                         & --                                                                           & --                                                                           & --                                                                           & --                                                                            & --                                                                           & --                                                                           & --                                                                            & --                                                                  & \begin{tabular}[c]{@{}c@{}}\textbf{-0.6037},\\\textbf{\textless{}0.001}\end{tabular}                       \\ 
\hline
{Number of links}                                                                   & --                         & --                                                                           & --                                                                           & --                                                                           & --                                                                            & --                                                                           & --                                                                           & --                                                                            & --                                                                  & --                                                                                      \\
\hline
\end{tabular}
\caption{\label{table: rp} Structural metrics at the VOI scale. Pearson correlation coefficient $r$ and corresponding $p$-values between the structural metrics (Fig. \ref{fig:multiscale_analysis}) are shown. In each cell, the upper value is $r$ and the lower value is $p$. Significant correlations with $p$ less than 0.05 are highlighted in bold.}
\end{table*}

\begin{figure}[hbp]
\centering
\includegraphics[width=\linewidth]{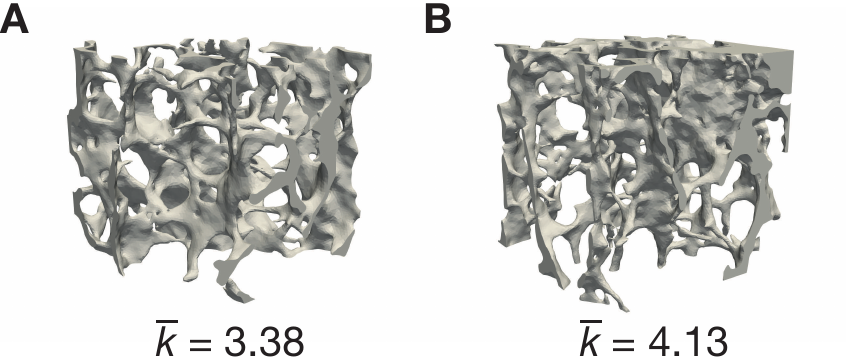}
\caption{\label{fig:node_degree_samples} {Illustration of 3-dimensional structure for two example VOIs. A: This sample corresponds to the example low-degree VOI indicated in Fig \ref{fig:multiscale_analysis}A and has an average node degree of 3.38, the smallest of any of the VOIs analyzed in this paper. B: This sample corresponds to the example high-degree VOI in Fig \ref{fig:multiscale_analysis}A and has an average node degree of 4.13, the largest node degree of all VOIs.}}
\end{figure}

Topological characteristics of nodes considered here include degree and weighted node degree. Degree refers to the number of links connected to a node, while weighted node degree is the sum of the weights of the links connected to a node. {For both of these measures, nodes of degree less than 3 are not considered, as the presence of these nodes is directly dependent on locations of the boundaries of VOIs. For example, nodes of degree 1 represent the ends of trabecular bone at the boundaries of VOIs. Nodes of degree 2, which are rare in trabecular bone and theoretically should not exist based on the definition of the trabecular networks, are the result of large ``chunk''-like pieces of bone, which are classified as nodes by the Skel2Graph algorithm, connected to two trabeculae.}

For links, we consider geometric properties relevant to spatially-embedded networks: average thickness (``trabecula width''), link length, vertical orientation (``Z-orientation'', $Zo$), and weighted vertical orientation (``weighted Z-orientation'', $Zo_w$). We define Z-orientation as the dot product of the position vector of a link with the unit vector in the Z-direction (superior-inferior direction). Z-orientation ranges from 0 to 1, where 0 and 1 refer to a link that is perpendicular and parallel to the Z-direction, respectively. Weighted Z-orientation is defined as the Z-orientation of a link multiplied by the corresponding weight of the link. We also analyze the average width of the pores (``pore width'') between trabeculae. Pore width is a metric we introduce to examine the distribution of spaces between trabecula at the smaller scale. Details regarding the calculation of these metrics are included in the Supplementary Information.

At the mesoscale, we compare averages of the above properties within each VOI across the vertebral body. Following bone histomorphometry conventions, the average width of trabeculae in a VOI is called trabecular thickness, or Tb.Th, and the average of pore widths in a VOI is called trabecular separation or trabecular spacing, abbreviated Tb.Sp (Supplementary Information) \cite{guidelines_assessment,bonej}. For weighted Z-orientation (but not unweighted orientation), we use the sum of the weighted Z-orientation of the links in a VOI as the VOI-scale measure, rather than the mean. We also determine the assortativity of each VOI, which is the tendency of nodes to be connected to other nodes that have similar properties. In this paper, we specifically determine degree assortativity, the tendency of nodes to be connected to nodes of similar degree. Nodes with a high assortativity (near 1) are said to display assortative mixing and nodes with low assortativity (near -1) are said to display disassortative mixing. Networks with assortativity near 0 are called neutral. Furthermore, we compute the volume fraction (BV/TV), a traditional histomorphometric quantity, and the total number of links in the network model of each VOI.

Fig. \ref{fig:multiscale_analysis}A-G compares within-VOI (left) and across-VOI (right) distributions for seven structural metrics (node degree, weighted node degree, pore width/Tb.Sp, trabecula width/Tb.Th, Z-orientation, weighted Z-orientation, and link length). The within-VOI plot shows distributions of each respective metric at the node/link scale for three representative VOIs that illustrate the within-VOI statistical distribution of the structural metric for representative high, medium, and low values of the corresponding VOI-scale average. The across-VOI plot illustrates the spatial distribution of the average of the metric in each VOI. Fig. \ref{fig:multiscale_analysis}H-J illustrates only across-VOI distributions for three metrics: number of links, assortativity, and volume fraction, which are not defined at the individual node/link scale. 

Distributions of node degree (Fig. \ref{fig:multiscale_analysis}A) within a VOI consistently demonstrate a peak at degree 3 and a tail extending to larger degree values. { In general, VOIs with a higher average degree and a higher peak at degree 3 also contain a few nodes of degree greater than 10. The yellow square-marked curve in Fig. \ref{fig:multiscale_analysis}A, corresponding to the largest node degree, is one example, containing nodes of degree 10, 11, 13, 15, 16, 22, 24, 37, 72, and 110. Nodes of degree greater than 20 are not shown so that the low degree behavior of the distributions is visible. These nodes are responsible for the yellow curve having the highest average node degree, despite the fact that the light blue diamond-marked curve has more nodes of degree 3 through 9.} Nodes of such high degree are uncommon in most of the trabecular bone samples analyzed in this paper. They tend to exist only in VOIs that contain dense regions of bone. These regions do not share the characteristic rod-like geometry of most trabecular bone, but are connected to many trabeculae due to their large surface area. In the network conversion process, these regions are approximated as nodes of unusually high degree.

{To illustrate how the 3-dimensional structure of VOIs varies for different values of average degree, Fig. \ref{fig:node_degree_samples} shows the continuum models generated by meshing the VOIs with the smallest (Fig. \ref{fig:node_degree_samples}A) and largest (Fig. \ref{fig:node_degree_samples}B) average node degrees. These correspond to the blue square-marked tile and the red circle-marked tile in the tilemap of Fig. \ref{fig:multiscale_analysis}A. Fig. \ref{fig:node_degree_samples}A displays a web-like structure throughout its volume with visible trabeculae. Fig. \ref{fig:node_degree_samples}B displays an example of a VOI that contains a node of incredibly high degree. This node is located in the upper left corner of the the figure, where the VOI contains a dense section of bone. }

The distributions of weighted node degree (Fig. \ref{fig:multiscale_analysis}B) consistently display peaks between 10 and 15 while having significantly smaller fraction of nodes of weighted degree less than 10 and greater than 20. Certain VOIs have significantly higher weighted node degree due to the presence of high degree nodes connected to links of large weight. These nodes can have weighted degrees in the hundreds, while the majority of nodes have weighted degrees less than one hundred. As a result, full distributions of the weighted node degree of VOIs are heavily right-skewed. In Fig. \ref{fig:multiscale_analysis}B, all three of the VOIs have such high weighted node degrees; we only show the nodes with weighted node degree between 0 and 60 so that the shape of each distribution is visible. Despite only showing a fraction of the full range of this plot, we only obscure about 0.2\% of the nodes in each of the three distributions, which make up the long tails of each of the distributions.

The trabecular spacing Tb.Sp (Fig. \ref{fig:multiscale_analysis}C) in a VOI varies greatly across the bone volume, ranging from 0.5 to 1.0 mm. Distributions of pore width within a VOI also vary in shape. The dark blue square-marked VOI, corresponding to the lowest Tb.Sp, exhibits a relatively symmetric distribution centered around 0.5 mm, while the distributions of the green diamond (mid-range Tb.Sp) and yellow (largest Tb.Sp) circle-marked VOIs are peaked at higher pore width values, with a heavy tail at low Tb.Sp. 

The trabecular thickness (Tb.Th) (Fig. \ref{fig:multiscale_analysis}D) of a VOI ranges from 0.12 to 0.35 mm, but the majority of VOIs have a Tb.Th less than 0.2 mm. The distributions of trabecula width within a VOI tend to have a sharp peak at small widths around 0.15 mm followed by a tail. The length of the tail reflects the size of the Tb.Th, with the dark blue square-marked distribution having the shortest tail and smallest Tb.Th.

The distributions of Z-orientation (Fig. \ref{fig:multiscale_analysis}E) indicate that some VOIs (e.g. the blue square-marked distribution with the smallest average Z-orientation) contain more trabeculae oriented perpendicular to the Z-axis, while others have more trabeculae oriented along the Z-axis (e.g., the yellow circle-marked distribution with the highest average Z-orientation). Overall, the mean Z-orientation does not vary greatly between the VOIs and ranges from 0.45 to 0.5, where the lower limit indicates VOIs that contain a slight prevalence of trabeculae oriented transverse to the Z-axis.

The distributions of weighted Z-orientation (Fig. \ref{fig:multiscale_analysis}G) consistently display a decay with increasing length. The VOI-scale color map illustrates the sum of all weighted Z-orientation values in each VOI, rather than the mean, in order to facilitate comparison with VOI-scale (unweighted) Z-orientation. While weighted Z-orientation at the link scale ranges from 0 to 2.2$\times 10^{-3}$, it ranges from 0.45 to 0.52 at the VOI scale. This narrow range can be attributed to our general observation that the sum of the thicknesses of the links in a VOI is usually roughly twice that of the sum of the thickness of each link multiplied by its Z-orientation. Thus, when dividing these quantities to get the weighted Z-orientation, we find values that are close to 0.5.

Distributions of link length (Fig. \ref{fig:multiscale_analysis}G) consistently demonstrate a large decaying behavior, with each VOI having hundreds of links of length about 0.2 mm but fewer than 20 links of length 0.9 mm or greater. The average link length of a VOI is heavily dependent on the range of lengths of the VOI. For instance, the yellow circle-marked VOI, which has the largest average link length of all VOIs, contains links as long as 1.7 mm, while the longest links in the blue square-marked VOI, which has the shortest average link length, are about 1.2 mm.

The number of links in the network model of each VOI varies greatly over the analyzed region (Fig. \ref{fig:multiscale_analysis}H). The network with the fewest links contains about 950 links, while the network with the greatest contains about 2600 links. However, a majority of networks contain fewer than 1500 links.

Fig. \ref{fig:multiscale_analysis}I shows that the VOIs analyzed in this paper display neutral mixing (assortativity near 0), with the assortativity values ranging from -0.08 to 0.12. This indicates that the nodes in these trabecular bone networks show no tendency to mix with nodes of either similar or dissimilar degree. 

The majority of the VOIs have a volume fraction less than 0.2 (Fig. \ref{fig:multiscale_analysis}J). Seven adjacent VOIs on the left side of the plot have slightly higher volume fraction, signifying a denser set of trabecular networks spanning that region.

Table \ref{table: rp} contains the Pearson correlation coefficients ($r$-values) and corresponding probability values ($p$-values) for each pair of structural metrics, with significant correlations highlighted in bold. We define a weak correlation as corresponding to the absolute value of $r$-values ranging from 0 to 0.3, moderate correlation as 0.3 to 0.6, and strong correlation as 0.6 to 1. 
We assert that there is strong evidence for a linear correlation (a correlation coefficient is significant) if $p \leq 0.05$. Assortativity and weighted node degree are significantly correlated with all of the other structural metrics. 
Volume fraction, Z-orientation, weighted Z-orientation,and link number are significantly correlated with all metrics except link length. Trabecular spacing is not significantly correlated with trabecular thickness or link length. Trabecular spacing is strongly negatively correlated with volume fraction, as is expected, and is also moderately correlated with Z-orientation. That is, a VOI with large average pore width tends to contain links that are less aligned with the vertical axis.  Trabecular thickness is moderately negatively correlated with Z-orientation and weighted Z-orientation. Hence, in our sample, VOIs with thicker trabeculae on average may tend to contain trabeculae less aligned with the vertical axis. 

\begin{figure}
\includegraphics[width=\linewidth]
{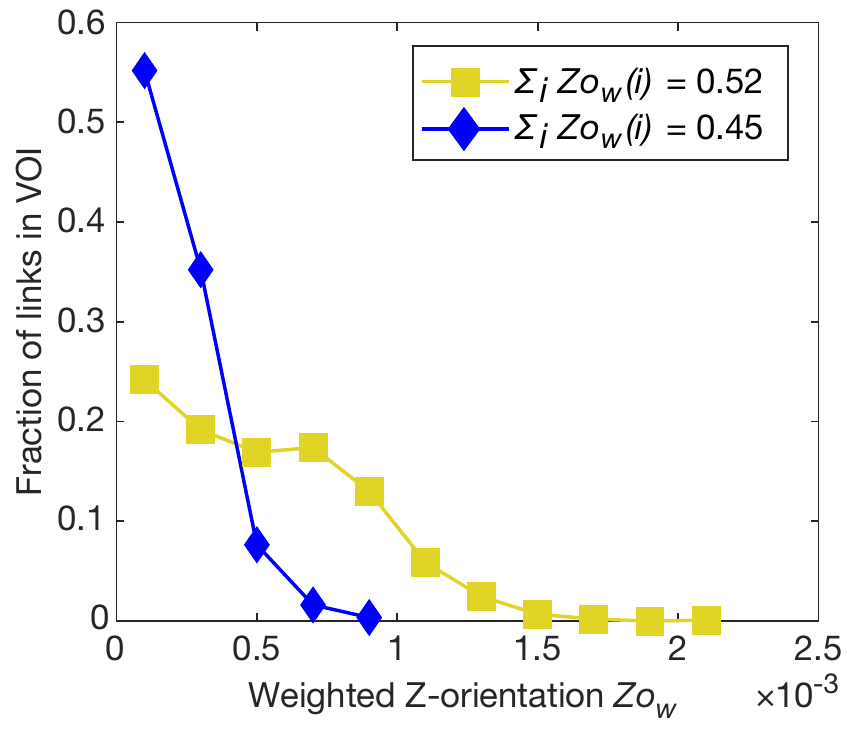}
\caption{\label{fig: wzo} Distributions of weighted Z-orientation illustrating differences between high average $Zo_w$ VOIs (yellow squares) and low average $Zo_w$ VOIs (blue diamonds). The distribution of the representative high $Zo_w$ VOI (marked in Fig. \ref{fig:multiscale_analysis}F by a circle) is much broader and displays a heavier tail compared to the narrower distribution of the representative low $Zo_w$ VOI (marked in Fig. \ref{fig:multiscale_analysis}F by a square).}
\end{figure}

Weighted Z-orientation is negatively correlated with both the number of links in a VOI and the volume fraction. Fig. \ref{fig:multiscale_analysis}F shows that the VOIs with high weighted Z-orientation are in the regions with the fewest links (Fig. \ref{fig:multiscale_analysis}H) and the lowest volume fraction  (Fig. \ref{fig:multiscale_analysis}J). We were initially surprised by this result. However, for the bone sample shown in Fig. \ref{fig:VOI_coarse_grain}, we observe that VOIs with lower volume fraction have a larger fraction of thicker links aligned with the Z-axis. Fig. \ref{fig: wzo} shows the distribution of weighted Z-orientations for the VOIs with the highest (yellow diamond-marked curve) and the least (blue diamond-marked curve) weighted Z-orientation. The yellow curve furthermore has among the fewest links of all VOIs and one of the lowest volume fractions, while the blue curve has among the most links and one of the largest volume fractions. The yellow distribution has a larger fraction of links with $Zo_w > 0.5$ than the blue distribution. We find that this is true for all VOIs with higher average weighted Z-orientation but low volume fraction and low number of links; they tend to have a larger range of weighted Z-orientation with a larger fraction of vertically oriented links. 

\begin{figure}[h]
\centering
\includegraphics[width=\linewidth]{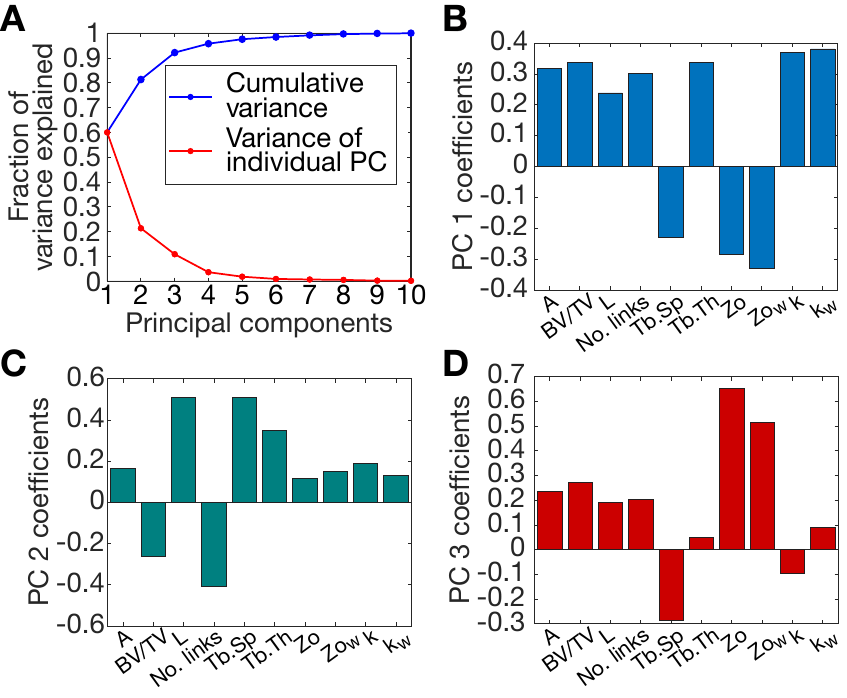}
\caption{\label{fig:pca} Principal component analysis of structural metrics. A: Fraction of variance explained by each principal component. The (upper) blue curve indicates the fraction of cumulative variance explained, while the (lower) red curve indicates the fractions explained by each principal component. The first three principal components explain 92.1\% of the total variance in the structural metric data. B-D: Correlation coefficients of the first three principal components of the structural metric data feature space. PC 1, which explains about 60\% of the data, only moderately or weakly correlates to any of the individual metrics. This is also true for PC 2, which explains about 21\% of the variance in the data. Z-orientation is strongly correlated to PC 3, which explains about 11\% of the variance in the data.}
\end{figure}

We use principal component analysis (PCA) to identify uncorrelated metrics that explain the majority of the variation in the VOI mesoscale structural data (Fig. 3). PCA was conducted using the Statistics and Machine Learning Toolbox for MATLAB. 
We examine the fraction of the total variance in structural metrics explained by each of the principal components (PCs) individually and cumulatively (Fig. \ref{fig:pca}A). 
The first PC explains approximately 60\% of the variance, while the second and third explain approximately 21\% and 11\% respectively. In total, they explain approximately 92\% of variance in the data. Since all the other components explain less than 10\% of the variance in the data, we focus further analysis on only the first three PCs.

Fig. \ref{fig:pca}B-D 
shows the correlation coefficients between the structural metrics and each of the first three PCs. PC 1 and PC 2 are either weakly or moderately correlated to all of the metrics.  Notably, all of the correlation coefficients in PC 1 are relatively similar (between 0.25 and 0.40). PC 3 is strongly correlated to Z-orientation and moderately correlated to node degree and weighted Z-orientation. PC 1 and PC 2 explain the majority of the variance in the structural metrics and are at best moderately correlated to the individual structural metrics. This result indicates that no smaller subset of the metrics (or linear combinations of them) can be used to capture the majority of the variance in the data, despite the significant correlations between almost all the structural metrics (Table \ref{table: rp}). \\
\subsection{\label{mechanicalnetworkanalysis} Finite element analysis}

To analyze mechanical response, we convert the bone networks into finite element models that consist of beam elements representing each link. We refer to these as ``beam models" (Fig. \ref{fig:samplepipeline}). We also construct continuum models generated from meshing the original {\textmu}CT images (Fig. \ref{fig:samplepipeline}) to serve as an \textit{in silico} validation of the beam models. We analyze both the bulk force-displacement response to compressive loading of the beam models and the distribution of stress in the beams. We individually carry out this analysis for each VOI in Fig. \ref{fig:VOI_coarse_grain}. Furthermore, we investigate how the structural properties of trabecular bone contribute to its mechanical response. We calculate the stiffness of the bone network in each VOI, and investigate correlations between the effective moduli and the structural metrics shown in Fig. \ref{fig:multiscale_analysis}.

We develop the beam models by converting each link in a network to a beam element (Fig. \ref{fig:samplepipeline}). {The beam elements are rigidly connected such that, under deformation, the angle between two beams remains the same.} The resulting models function as 3-D realizations of the network model. Simulations with the continuum models, which are full-scale mesh reconstructions, are used to validate the simulation results of the beam model. The beam-element and continuum models are analyzed in Abaqus FEA (Dassault Syst\`{e}mes, V\'{e}lizy-Villacoublay, France). Compared to the continuum models, the beam models correspond to a reduction in the degrees of freedom by about one order of magnitude, and require about an order of magnitude less computation time to solve.

We simulate compressive (top to bottom) loading in the linear-elastic regime. The elastic modulus of each beam is equal and set to 10 GPa, and the Poisson ratio is set to 0.16, following ranges reported for trabecular bone in the literature \cite{rho1993_modulus,poissonRatiosource}. However, since the analysis is linear-elastic, a different choice of values simply corresponds to a linear scaling of the results.

The topmost and bottommost nodes of the VOI are identified as those lying in the transverse planes on the top and bottom of the VOI. The bottom nodes are held fixed in all dimensions, while the top nodes are displaced slowly in the -Z (superior-inferior) direction at a constant loading rate.

\begin{figure}
\centering
\includegraphics[width=\linewidth]{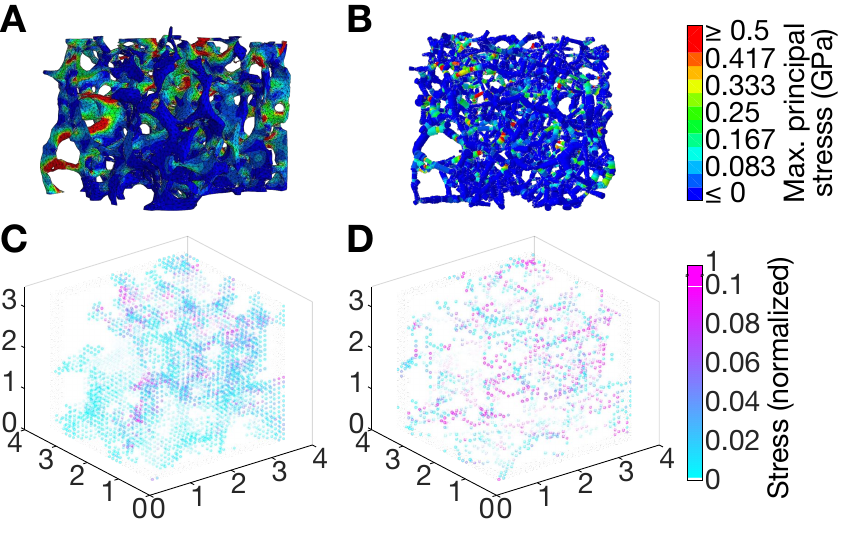}
\caption{\label{fig: beam_cont_compare} Finite element models of trabecular bone, for a sample VOI. The continuum model (A) and beam-element model (B) generated from the same VOI are compressed from the top. Colors show maximum principal stress in each element at the end of the simulation. C-D: Coarse-grained spatial distributions of maximum principal stress for the continuum (C) and beam-element (D) models. Each model is divided into a regular grid of (0.11 mm)$^3$ bins; each point corresponds to the average stress in one bin. Stress is normalized to the highest stress value (measured for a single element) in each model.}
\end{figure}

For each VOI, we validate the beam model by comparing results of the simulated compression with that of the continuum model, using the continuum result as an \textit{in silico} validation. In the linear-elastic regime, initial comparisons (not shown) of the force-displacement curves indicate that the beam model has lower stiffness compared to the continuum model. In order to match the stiffness of the beam model to the continuum model, the radius of each beam was increased. For the example VOI analyzed in Fig. \ref{fig: beam_cont_compare}, an overall scale factor of 1.55 was required to match the force-displacement response (Supplementary Information). Our use of the scale factor is attributed to geometric differences between the beam and the continuum models. The cross-section of a trabecula is not exactly circular, but is approximated as circular in the formulation of the beams in the finite element model. Using a square cross-section for the beams while keeping the same thickness increases the overall cross-sectional area of the model and would slightly reduce but not entirely eliminate the scale factor. Moreover, while the individual beams have uniform thickness, the continuum model trabeculae have inhomogeneous thickness. Additionally, while the beam model approximates the branch points as nodes, the branch points in the continuum model are regions of bone with significant bulk properties that add to the stiffness of trabecular bone. The models used to produce the results shown in this paper contain beams with circular cross-sections.

Fig. \ref{fig: beam_cont_compare}A illustrates the stress states of each element of the beam and continuum models at the end of the simulations of linear compressive loading, colored according to the stress in each element. Stress in this paper refers specifically to maximum principal stress, the first (diagonal) element of the stress tensor in a coordinate system with no shear stress. Because the two models have different numbers of elements and different types of elements (beams in the beam model, tetrahedral elements in the continuum model), to facilitate a comparison of the spatial stress distribution, we coarse-grain each model by dividing the (3.7 mm)$^3$ VOI into a regular grid of (0.185 mm)$^3$ bins and average the stress in each bin (Fig. \ref{fig: beam_cont_compare}B). While the locations of high stress are similar, the highest stresses in the continuum model are almost an order of magnitude greater than the beam model. (Note that Fig. \ref{fig: beam_cont_compare}B plots stress normalized by the maximum stress in one individual element for each model.) Both models exhibit a low-to-high-stress gradient along the +Z (superior-inferior) direction. However, this gradient is more pronounced for the beam model, while the continuum model contains greater spatial variation in stress. A trabecula is typically non-uniform in thickness, and can contain significantly thinner regions, but the network conversion process averages the thickness over a trabecula to produce the beam model. Hence, the continuum model can contain much thinner regions than the beam model, as well as relatively sharp corners that are smoothed in the beam model but which could be regions of localized stress in the continuum model.

\begin{figure*}
\includegraphics[width=\linewidth]{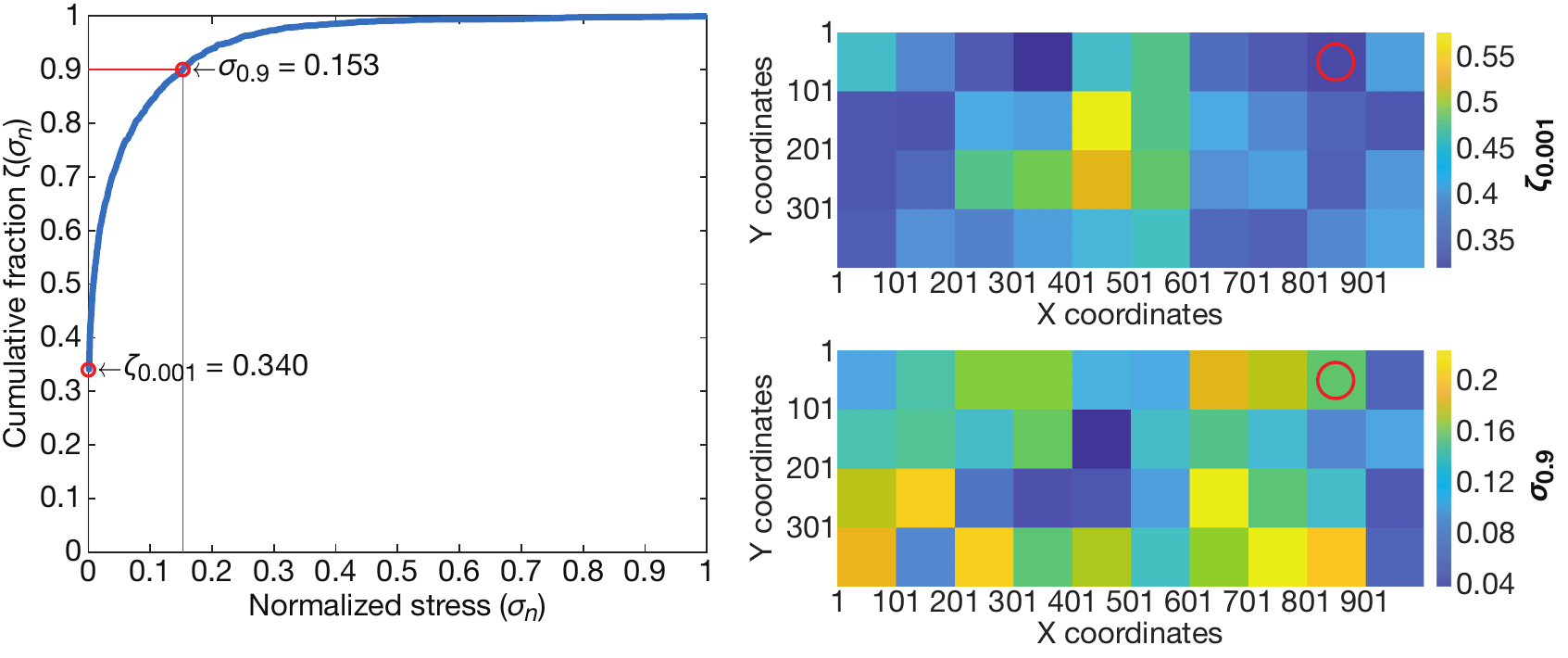}
\caption{\label{fig: stressCDF} Distributions of stress. Left: Distribution of normalized maximum principal stress in the beam elements of an example beam model (Fig. \ref{fig: beam_cont_compare}) under compressive loading in the linear regime. To calculate normalized stress ${\sigma}_n$, we normalize stress ${\sigma}$ with the largest value of stress in a beam at the final timestep of the compressive loading simulation. The function $\zeta$ is the fraction of the beams that bear a normalized stress less than or equal to ${\sigma}_n$. $\zeta_{0.001}$ is defined as the fraction of beams that bear a normalized stress less than or equal to 0.001 and $\sigma_{0.9}$ is defined as the normalized stress that satisfies the equation $\zeta (\sigma_{0.9})=0.9$. In this VOI, 34\% of the beams bear a normalized stress less than 0.001 ($\zeta_{0.001} = 0.340$), while 90\% of beams bear a normalized stress less than or equal to 0.153 ($\sigma_{0.9}=0.153$). Right: Spatial distributions of $\zeta_{0.001}$ and $\sigma_{0.9}$ across the sample. The example VOI is indicated by the red circle.}
\end{figure*}

	During loading, the stress carried by individual beams in the VOI varies significantly. Fig. \ref{fig: stressCDF} shows the distribution of normalized stress in the beam model sample (Fig. \ref{fig: beam_cont_compare}A) undergoing compressive loading in the linear regime. While Fig. \ref{fig: stressCDF} shows the distribution of stress during the final timestep of loading, the shape of this distribution remains constant in the linear regime for all timesteps.  We define the parameters $\zeta_{0.001}$ and $\sigma_{0.9}$ to characterize this distribution, where $\zeta_{0.001}$ is the fraction of beams with normalized stress less than or equal to 0.001, and where ninety percent of beams bear a stress less than or equal to $\sigma_{0.9}$. In the VOI shown in Fig. \ref{fig: stressCDF}, $\zeta_{0.001}=$ 0.340 and $\sigma_{0.9}=$ 0.153. For the VOIs studied in this paper, the average value of $\zeta_{0.001}$ is 0.410, while the average value of $\sigma_{0.9}=$ 0.136. $\zeta_{0.001}$ ranges from 0.3 to 0.6 and $\sigma_{0.9}$ ranges from 0.038 to 0.224 across all VOIs.

\begin{figure}
\includegraphics[width=\linewidth]
{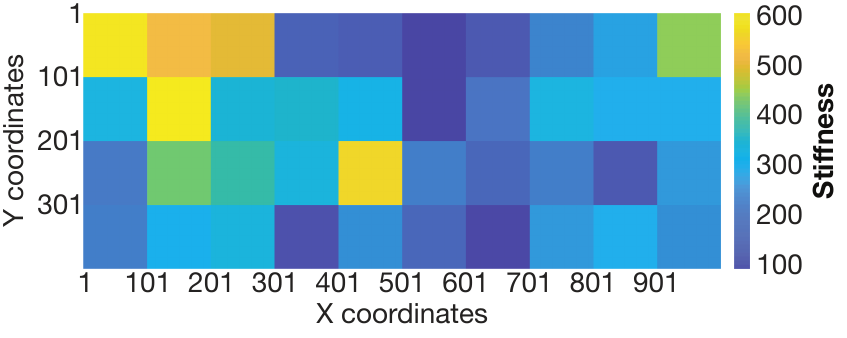}
\caption{\label{fig: stiff} Spatial distribution of stiffness across the vertebral body. The color of each tile represents the stiffness for one VOI.}
\end{figure}

\subsection{Relating structure and mechanics}

We investigate the relationship between structural properties -- histomorphometric, geometric, and network-topological metrics -- and mechanical properties at both the individual link (or node) scale and the VOI mesoscale.

\subsubsection{Individual Link Scale}

At the scale of individual links, we analyze the stress borne by each link during the final timestep of compression in our simulations. We compare link structural features to the distribution of stresses about the links to determine whether any structural properties are correlated to mechanical properties at the individual link scale. For each VOI, we calculate Pearson correlation coefficients and corresponding $p$-values between the stresses and link metrics. We present the average of the $r$ and $p$-values across all VOIs in Table \ref{table: rp_small_scale}. 
We observe significant but weak correlations between stress and Z-orientation. Weighted Z-orientation is also significantly correlated with both Z-orientation and trabecula width, which is unsurprising due to its definition. Additionally, the length of the links is weakly, negatively correlated to Z-orientation.

\begin{table}[tbp]
\centering
\def\arraystretch{1.5}
\setlength\tabcolsep{2pt}
\begin{tabular}{|c||c|c|c|c|c|} 
\hline
Metric                                                        & \multicolumn{1}{l|}{$S$} & Width                                                         & $L$                                                        & $Zo$                                                                                    & $Zo_w$                                                                        \\ 
\hline\hline
Stress ($S$)                                                    & --                      & \begin{tabular}[c]{@{}c@{}}-0.0331,\\ 0.2810 \end{tabular} & \begin{tabular}[c]{@{}c@{}}0.0453,\\ 0.2106\end{tabular} & \begin{tabular}[c]{@{}c@{}}\textbf{-0.1320},\\ \textbf{ 0.0081} \end{tabular} & \begin{tabular}[c]{@{}c@{}} -0.0013,\\ 0.3641\end{tabular}   \\ 
\hline
\begin{tabular}[c]{@{}c@{}}Trabecula\\width \end{tabular} & --                      & --                                                          & \begin{tabular}[c]{@{}c@{}}0.0294,\\0.3408\end{tabular}  & \begin{tabular}[c]{@{}c@{}}-0.0868,\\ 0.0797 \end{tabular}                              & \begin{tabular}[c]{@{}c@{}}\textbf{0.5137},\\\textbf{\textless{} 0.001}\end{tabular}   \\ 
\hline
Link length ($L$)                                               & --                      & --                                                          & --                                                        & \begin{tabular}[c]{@{}c@{}} \textbf{-0.0792},\\\textbf{0.0462} \end{tabular} & \begin{tabular}[c]{@{}c@{}}-0.0187,\\0.2003\end{tabular}  \\ 
\hline
Z-orientation ($Zo$)                                           & --                      & --                                                          & --                                                        & --                                                                                      & \begin{tabular}[c]{@{}c@{}}\textbf{-0.4985},\\\textbf{\textless{} 0.001}\end{tabular}   \\ 
\hline
\begin{tabular}[c]{@{}c@{}}Weighted\\Z-orientation ($Zo_w$) \end{tabular}                   & --                      & --                                                          & --                                                        & --                                                                                      & --                                                                             \\
\hline
\end{tabular}
\caption{\label{table: rp_small_scale}Comparing stress with structural metrics at the individual link scale. Pearson correlation coefficient $r$ and corresponding $p$-values between structural metrics and stress at the individual link scale are shown. In each cell, the upper value is $r$ and the lower value is $p$. The values reported here are the averages of the coefficients over all VOIs. Significant correlations with $p$ less than 0.05 are highlighted in bold.}
\end{table}

\subsubsection{VOI Scale}
\begin{figure*} [h!]
\includegraphics[width=0.65\textwidth]{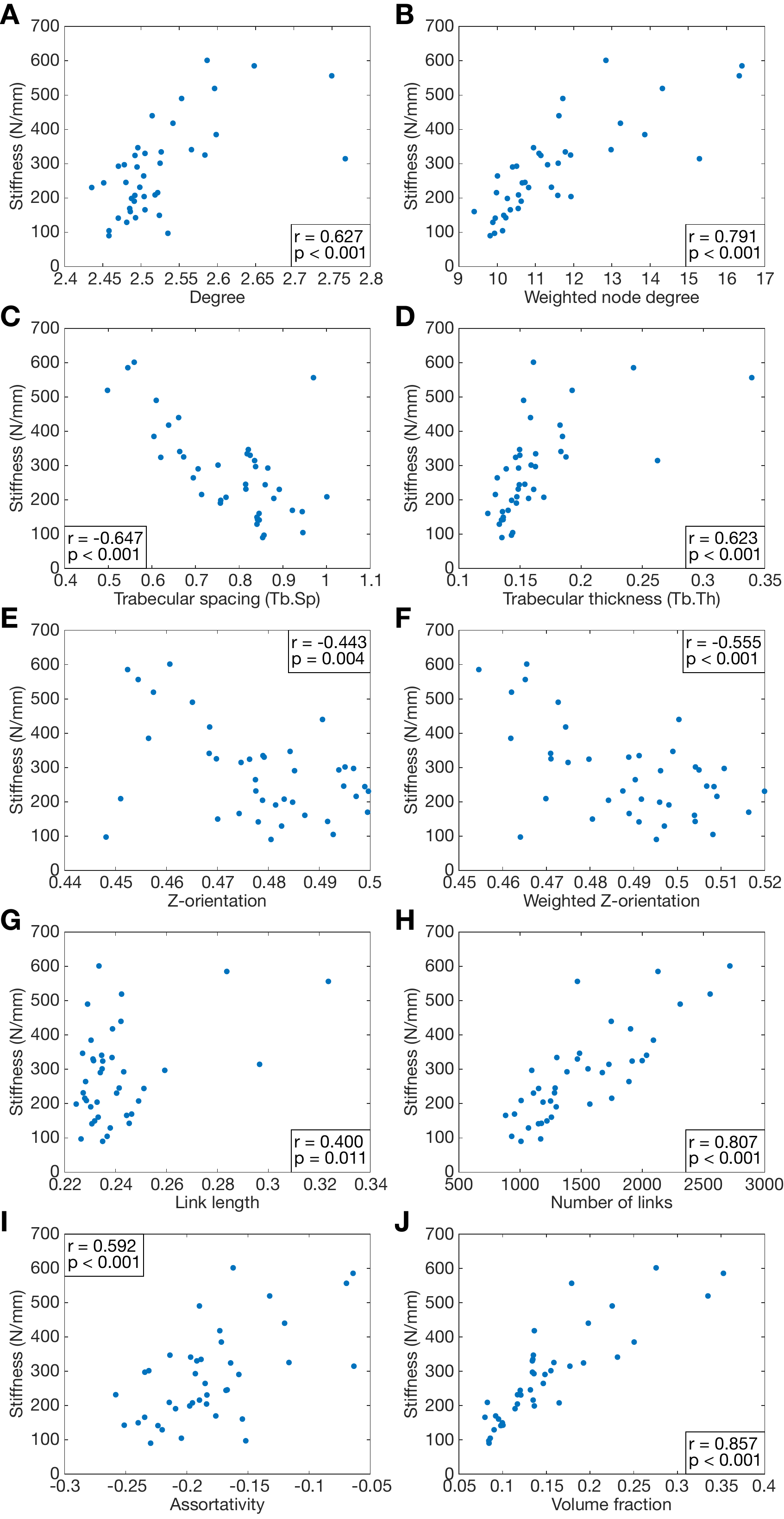}
\caption{\label{fig:bga_modulus_scale_correlation_analysis} Stiffness compared with structural metrics at the VOI scale. For each VOI, the stiffness is plotted against the average degree (A), weighted node degree (B), trabecular spacing (C), trabecular thickness (D), Z-orientation (E), weighted Z-orientation (F), and link length (G), as well as the overall number of links (H), assortativity (I), and volume fraction (J).}
\end{figure*}

At the VOI mesoscale, we analyze the stiffness of each VOI. Stiffness is defined as the slope of the force-displacement curve in the linear regime. 
Fig. \ref{fig: stiff} shows the spatial distribution of stiffness across all VOIs. In the linear-elastic regime, the stiffness is a constant over the loading process for each individual VOI. Fig. \ref{fig:bga_modulus_scale_correlation_analysis} compares stiffness with ten network-topological, geometric, and traditional histomorphometric metrics. 

We find significant linear correlations between the stiffness of each sample and all structural metrics shown in Fig. \ref{fig:multiscale_analysis}. Stiffness is most strongly correlated with volume fraction ($r = 0.857$, $p < 0.001$), number of links ($r = 0.807$, $p < 0.001$), and weighted node degree ($r=0.791$, $p< 0.001$. We also observe significant, strong positive linear correlations between stiffness and degree ($r = 0.627$, $p < 0.001$) as well as stiffness and Tb.Th ($r = 0.623$, $p < 0.001$). Stiffness exhibits a significant, strong negative linear correlation with Tb.Sp ($r = -0.647$, $p < 0.001$). We also observe moderate but significant correlations between stiffness and assortativity ($r = 0.592$, $p < 0.001$), link length ($r = 0.400$, $p = 0.011$), Z-orientation ($r =-0.443$, $p = 0.004$), and weighted Z-orientation ($r = -0.555$, $p < 0.001$). These results indicate that the number of links, degree, weighted degree, and assortativity can be informative network topological features to supplement BMD in characterizing bone strength. Furthermore, weighted Z-orientation can be an informative geometric property of the spatially-embedded network, in addition to volume fraction, trabecular spacing, and trabecular thickness for histomorphometric analysis. The strong correlation between stiffness and volume fraction shows that trabecular networks tend to be stiffer as the ratio of bone volume to pore volume increases, and the strong correlation between stiffness and weighted node degree indicates that stiffer trabecular networks have larger numbers of thicker trabecula connected to each other. 

To determine whether all ten metrics are necessary to predict stiffness, we performed a multiple linear regression using the following model:

\begin{center}
\begin{equation}
\label{eqn: regressionmodel} y = \beta_0 + \beta_1 x_1 + \beta_2 x_2 + \ldots + \beta_n x_n,
\end{equation}
\end{center}

where $y$ corresponds to stiffness and the $x_i$ correspond to each of the structural metrics ($n$ = 10). The data is standardized prior to fitting, and the  standardized linear coefficients $\beta_i$ are listed in Table \ref{table: regression_coefficients}.   

\begin{table}
{
\def\arraystretch{2}
\setlength\tabcolsep{1pt}
\begin{tabular}{|c||c|c|c|}
\hline
                & $\beta_i$ (fit) & Standard error & $p$   \\
\hline Intercept       & 0  & 0.057         & 1  \\
\hline Assortativity~  & 0.009  & 0.097         & 0.927  \\
\hline \textbf{Degree}          & \textbf{-0.635}  & \textbf{0.191}         & \textbf{0.002}  \\
\hline \textbf{Weighted degree} & \textbf{0.872}   & \textbf{0.295}           & \textbf{0.005}  \\
\hline Volume fraction & 0.872  & 0.215         & 0.59  \\
\hline \textbf{Tb.Sp}           & \textbf{0.416}   & \textbf{0.192}         & \textbf{0.038}  \\
\hline Tb.Th           & -0.172 & 0.277         & 0.053 \\
\hline \textbf{Link length}     & \textbf{-0.699}   & \textbf{0.299}         & \textbf{0.026}  \\
\hline \textbf{Z-orientation}   & \textbf{0.753}  & \textbf{0.355}           & \textbf{0.042}  \\
\hline \textbf{Weighted Z-orientation}    & \textbf{1.30}   & \textbf{0.230}         & \textbf{\textless{} 0.001}  \\
\hline \textbf{Number of links} & \textbf{0.914}  & \textbf{0.287}       & \textbf{0.004} \\
\hline
\end{tabular}}
\caption{\label{table: regression_coefficients} Standardized linear coefficients, standard errors, and $p$-values for a multiple linear regression model relating stiffness with the ten structural metrics. Metrics with $p < 0.05$ are highlighted.}
\end{table}

There are forty different observations included in the regression analysis, one for each VOI. These observations correspond to the ten metrics calculated for each VOI.
We find that the linear model is a significantly better fit to the data ($p < 0.001$) than a constant model under the $F$-test, and that the ten metrics are strongly predictive of stiffness {(coefficient of determination $r^2$ = 0.905). Furthermore, significance values for each of the individual metrics (Table \ref{table: regression_coefficients}) indicate the significant contribution of seven metrics to the prediction of stiffness: degree, weighted node degree, trabecular spacing, link length, Z-orientation, weighted Z-orientation, and the number of links.} Most notably, volume fraction does not contribute significantly in the linear model, despite its strong correlation with stiffness. Furthermore, removing any of these {seven} metrics, as well as Tb.Th, from the model decreases the adjusted $r^2$ (Supplemental Material), which penalizes the number of explanatory variables in the model. For a linear model containing only the aforementioned {seven} significant metrics, adding any additional variable to the model also decreases the adjusted $r^2$. {This indicates that these seven metrics are the most informative metrics in predicting stiffness with a multilinear model. The model with ten significant metrics has an adjusted $r^2 = 0.872$. The model with seven significant metrics has an adjusted $r^2 = 0.882$.}

{We also performed a multiple linear regression using the model described by Eq. (\ref{eqn: regressionmodel}), with the principal components shown in Fig. \ref{fig:pca}. A model including all ten principal components has the same $r^2$ and adjusted $r^2$  as the model shown in Table \ref{table: regression_coefficients}. A model including the principal components which contribute most significantly to the prediction of stiffness ($p < 0.05$) has an $r^2$ of 0.878 and an adjusted $r^2$ of 0.864. These values indicate that a model consisting of the significant principal components does not perform as well as a model with the significant structural metrics.}

 \section{\label{sec:conclusions} Conclusion}
We introduce a network characterization of bone that provides a new framework for analyzing bone architecture. This approach incorporates existing mathematical and computational methods developed for graph theory and network science with finite element analysis, directly relating topological, geometric, and mechanical properties of trabecular bone. Moreover, the beam models developed in this paper provide streamlined, efficient alternatives to traditional methods of mechanical analysis of bone, which depend on computationally expensive image processing methods to conduct structural and finite element analysis. In this paper, we use the network characterization and beam models to analyze bone structure and mechanics at the scale of individual trabeculae and at the scale of 50 mm$^3$ volumes of interest. Further studies will involve investigating trabecular bone at larger scales, extending to the entirety of the vertebra. 
	
Our method of generating beam models of trabecular bone through skeletonization has some similarities with the network representation of soil samples developed in \cite{soil_networks}, as well as 3-D Line Skeleton Graph Analysis (LSGA) developed specifically for trabecular bone in \cite{pothuaud_development_lsga}. LSGA analyzes the mechanical properties of trabecular bone by skeletonizing bone images and converting the skeletons into FEM beam models. LSGA also achieves an improved force-displacement curve fit with a more detailed bone model only after the thickness of the beams is increased. However, unlike the LSGA method, we use network science methods to further analyze the bone topology in addition to creating beam models. Additionally, we use the network and beam models to characterize trabecular bone not only at the VOI scale, but also at the scale of individual trabeculae, which is not analyzed with LSGA.

We analyzed the within-VOI distribution of network-topological, geometric, and traditional histomorphometric properties at the sub-millimeter scale (the level of individual links), as well as the spatial distribution across VOIs at the millimeter scale. {While it would be informative to quantify the distributions of the individual metrics in detail, such an analysis is outside the scope of this paper. The goal of the structural analysis was to determine a set of useful metrics for describing trabecular bone structure, and we determined that these metrics are correlated with the stiffness of trabecular bone using 40 healthy trabecular bone samples. Future work can include a more comprehensive statistical analysis using a significantly larger data set containing both healthy and osteoporotic bone in order to characterize how the structural metrics vary as the overall health of bone samples decreases.}

{Though the distributions of structural metrics shown in Fig. 3 are generally not normal distributions, we determine the mean value of the metrics as a convenient statistic to differentiate between VOIs. We follow the convention in network science of characterizing a network by its mean degree and/or weighted degree, and the convention in histomorphometry of using the mean trabecular thickness, spacing, and length. Future work can involve a more extensive statistical analysis on a larger data set to identify other markers for characterizing trabecular architecture.}

Using principal component analysis, we find no subset of properties that captures the majority of the variation in the structural metrics, indicating that all metrics provide unique information about the structure of the trabecular networks. We also determine the Pearson correlation coefficients between structural metrics and stiffness, and find that stiffness is significantly (positively or negatively) correlated with all structural metrics analyzed. 
The strongest positive correlation observed was between stiffness and volume fraction, corroborating previous studies which also find that volume fraction explains a large percentage of the variation of stiffness in osteoporotic bone for similarly sized VOIs (spatial dimensions on the millimeter scale) \cite{elastic_properties_fabric_tensor,volume_fraction}. We furthermore demonstrate a positive correlation between stiffness and weighted node degree that is considerably stronger than the correlation between stiffness and degree (which characterizes connectivity without taking thickness into consideration) and the correlation between stiffness and trabecular thickness (which characterizes thickness without connectivity). This may indicate that stiffer networks contain links that are both thicker and more interconnected.

We use multiple linear regression to identify {seven} metrics that contribute the most to explaining the variance in the data in a linear regression model: {degree, weighted node degree, trabecular spacing, link length, Z-orientation, weighted Z-orientation, and the number of links} all had significant $p$-values in the multiple linear regression (Table \ref{table: regression_coefficients}). These metrics are determined by computing a slightly different set of measures: {node degree, trabecular thickness, trabecular spacing, link length, Z-orientation, and number of links. Additionally, we use multiple linear regression with the principal components of the structural metric data to determine whether or not they present a better fit to stiffness. We find that that a model consisting of the significant structural metrics fits the stiffness data better than a model consisting of the significant principal components (adjusted $r^2 = 0.882$ for the former versus adjusted $r^2 = 0.864$ for the latter).}

It is surprising that the analysis did not identify volume fraction as a significant ($p < 0.05$) variable for the prediction of stiffness, considering that volume fraction exhibits the strongest linear correlation with stiffness out of all 10 structural metrics in a linear regression model (Fig. \ref{fig:bga_modulus_scale_correlation_analysis}). This does not indicate that volume fraction is uninformative in the prediction of stiffness. Its lack of significance in the multiple linear regression implies that it does not improve the predictive ability of a linear model which includes the {seven} significant metrics previously indicated. However, volume fraction is known to be the primary predictor of stiffness in porous media \cite{cellular_materials}. Previous studies have indicated nonlinear relationships between mechanical properties, including compressive yield strength and elastic modulus, and volume fraction in trabecular bone \cite{volume_ash_fraction}.
In this paper, we use multiple linear regression analysis to identify the smallest subset of metrics that captures the most variation in stiffness; future work will extend the current regression model to account for the possible nonlinear dependence of stiffness on volume fraction and other variables in order to improve predictive power.

From the stress distribution across the elements of the beam models, we find that only a small number of beams withstand a load comparable to the maximal stress on a network, while the majority of links bear a stress less than or equal to one-third of this maximal stress. Further development of our modeling framework will extend the beam model to the nonlinear plastic regime, and ultimately to the point of failure, to investigate how the failure of individual links affects the distribution of stress on the network and the overall compressive strength of the network. This may prove informative in predicting the fracture susceptibility of a trabecular network and can serve as a biologically-motivated application of previous studies characterizing the failure of disordered elastic networks \cite{driscoll}. Furthermore, simulating the response of bone to other types of loading conditions, such as shearing, tension, or rapid impacts, can be useful in developing a comprehensive model of fracture.

Trabecular bone exhibits hierarchical organization at various scales. Individual trabeculae are made up of lamellae, which themselves are composed of mineralized collagen fibrils (MCFs, the ``building blocks'' of bone). MCFs consist of mineral plates embedded within a collagen matrix, and the microscale and nanoscale mechanics of MCFs contribute to overall bone elasticity \cite{jagerfratzl,hellmich2004,fritsch2009}. Future work can integrate results from different scales to provide a more complete characterization of bone from its molecular constituents to its architecture at large.

Our results identifying relationships between structural metrics and mechanical properties suggest these mesoscale metrics may prove informative for bone health. Extensions of our work to comparisons between healthy and osteoporotic bone samples may inform future diagnostics. In particular, extensions of the analyses of Table \ref{table: rp} and Fig. \ref{fig:bga_modulus_scale_correlation_analysis} to diseased bone may inform the characterization of fracture resistance by identifying structural differences between healthy and diseased bone. Additionally, applying network analysis to bone at various stages of disease or aging may provide insight into how healthy bone changes over time.

In clinical applications, high-resolution \textit{in vivo} measurements are increasingly appreciated as necessary for the evaluation of bone fragility. Innovative techniques for high-resolution data acquisition of fine tissue structure are already in development \cite{bioprotonics}, as well as techniques for \textit{in vivo} mechanical assessment such as reference point indentation \cite{rpi}. The methods developed in this paper aim to complement advances in medical diagnostic measurements by identifying biomarkers that may be useful to target using clinical procedures. Moreover, should high-resolution \textit{in vivo} imaging of human bone throughout the body become feasible, network models can be generated from bone scans of patients and used to assess fracture risk. Our framework can hence inform the development of improved procedures for assessing bone health and detecting the onset of disease.

\section{\label{sec: acknowledgements} Acknowledgements}
This work was supported by the Worster Summer Research Fellowship, the David and Lucile Packard Foundation, the Institute of Collaborative Biotechnologies through Army Research Office grant W911NF-09-D-0001, and the National Science Foundation under grants EAR-1345074 and CMMI-1435920. The content of the information does not necessarily reflect the position or the policy of the Government, and no official endorsement should be inferred.

\end{document}